\documentclass[10pt,conference]{IEEEtran}
\IEEEoverridecommandlockouts

\usepackage{graphicx}
\usepackage{stfloats}
\usepackage{textcomp}
\usepackage{xcolor}
\usepackage{xspace}
\usepackage[nointegrals]{wasysym}
\usepackage{pifont}
\usepackage{multirow}
\usepackage{flushend}
\usepackage{changes}
\usepackage{caption}
\usepackage{booktabs}
\usepackage{hyperref}
\usepackage{algorithm}
\usepackage{pdfpages}
\usepackage{amssymb}
\usepackage{tabularx}
\usepackage{multicol} 
\hypersetup{
colorlinks = true, 
linkcolor=black, 
citecolor=black, 
urlcolor=black 
}
\usepackage{url}
\usepackage{enumitem}
\usepackage{braket}
\usepackage{amsmath}
\usepackage[breakable]{tcolorbox}
\usepackage{colortbl}
\usepackage{pifont}
\usepackage{float}                  
\usepackage{subfig}                 
\usepackage{overpic}                
\usepackage{algpseudocode}

\usepackage{tikz}
\usetikzlibrary{quantikz}
\usetikzlibrary{positioning}
\usetikzlibrary{arrows.meta}
\usepackage{color}
\usepackage{listings}

\usepackage{changes} 
\makeatletter
\AddToHook{cmd/deleted/before}{\def\Changes@AuthorColor{red}} 
\usepackage{listings}
\usepackage{newfloat}
\usepackage{booktabs}    
\usepackage{tabularx}    
\makeatother

\renewcommand{\deleted}[1]{}

\DeclareFloatingEnvironment[
    fileext=loe,          
    listname=List of Examples, 
    name=Example,         
    placement=tbhp,
]{example}

\definecolor{bleudefrance}{rgb}{0.19, 0.55, 0.91}
\usepackage{times}
\usepackage{graphicx}
\usepackage{epsf}
\usepackage{verbatim}
\usepackage{url}
\usepackage{color}
\usepackage{alltt}

\newcommand{\SmallSpace}{\vspace*{-1.4ex}}

\newcommand\black[1]{\textcolor{black}{#1}}
\begin{document}
%
\title{QuanBench: Benchmarking Quantum Code Generation with Large Language Models}

\author{\IEEEauthorblockN{Xiaoyu Guo$\dagger$}
\IEEEauthorblockA{
\textit{Kyushu University}\\
guo.xiaoyu.961@s.kyushu-u.ac.jp}
\and
\IEEEauthorblockN{Minggu Wang$\dagger$}
\IEEEauthorblockA{
\textit{Kyushu University}\\
wang.minggu.065@s.kyushu-u.ac.jp}
\and
\IEEEauthorblockN{Jianjun Zhao*}
\IEEEauthorblockA{
\textit{Kyushu University}\\
zhao@ait.kyushu-u.ac.jp}
\thanks{$\dagger$ These authors contributed equally to this work.}
\thanks{* Corresponding author.}
}

\maketitle
\begin{abstract}
Large language models (LLMs) have demonstrated good performance in general code generation; however, their capabilities in quantum code generation remain insufficiently studied. This paper presents QuanBench, a benchmark for evaluating LLMs on quantum code generation. QuanBench includes 44 programming tasks that cover quantum algorithms, state preparation, gate decomposition, and quantum machine learning. Each task has an executable canonical solution and is evaluated by functional correctness (Pass@K) and quantum semantic equivalence (Process Fidelity). We evaluate several recent LLMs, including general-purpose and code-specialized models. The results show that current LLMs have limited capability in generating the correct quantum code, with overall accuracy below 40\% and frequent semantic errors. We also analyze common failure cases, such as outdated API usage, circuit construction errors, and incorrect algorithm logic. QuanBench provides a basis for future work on improving quantum code generation with LLMs.
\end{abstract}






\section{Introduction}
Recent advances in LLMs have significantly advanced software engineering tasks such as code generation~\cite{huynh2025large, athiwaratkun2022multi}, synthesis~\cite{austin2021program}, and completion~\cite{guo2023longcoder, lu2022reacc, wang2021code}. Transformer-based models such as ChatGPT~\cite{achiam2023gpt}, DeepSeek~\cite{deepseekai2024deepseekv3technicalreport}, Gemini~\cite{deepmind_gemini}, and LLaMA~\cite{touvron2023llama} demonstrate strong performance on general-purpose programming benchmarks, including HumanEval~\cite{chen2021evaluating}, MBPP~\cite{austin2021program}, and DS-1000~\cite{lai2023ds}, enabling the generation of standalone functions directly from natural language prompts. These benchmarks have supported the rapid adoption of LLMs in classical software development workflows.

While these models have achieved strong results on classical programming tasks, their capabilities in specialized domains such as quantum programming remain underexplored. Quantum programming differs from classical programming in both syntax and semantics~\cite{ying2016foundations,gay2006quantum}. It operates on quantum bits (qubits) rather than classical bits, involves unitary and reversible transformations, and depends on linear algebra and measurement-based computation~\cite{nielsen2010quantum}. These characteristics present unique challenges for LLM-based code generation, including the generation of semantically valid quantum circuits, the preparation of entangled states, gate decomposition, and the correct implementation of quantum algorithms. 

Although Qiskit HumanEval~\cite{vishwakarma2024qiskit} has been proposed to evaluate quantum code generation by adapting Qiskit-specific APIs, its focus remains largely on API compliance rather than on testing algorithmic reasoning or quantum semantic correctness. Currently, no benchmark systematically evaluates whether LLMs can generate functionally correct and semantically meaningful quantum programs from natural language descriptions, which is essential for practical quantum software development.

To address this gap, we present QuanBench, a benchmark designed to evaluate LLMs on quantum code generation tasks. Built on the Qiskit framework (version 0.46.0), QuanBench comprises 44 carefully curated tasks, each accompanied by a natural language prompt, a canonical implementation, and unit tests. The benchmark covers four categories of quantum programming tasks: quantum algorithm implementation (e.g., Grover’s algorithm \cite{long2001grover}, Quantum Fourier Transform \cite{weinstein2001implementation}), quantum state preparation (e.g., Bell and GHZ states), gate decomposition, and quantum machine learning (e.g., variational circuits \cite{cerezo2021variational}).

We evaluate model outputs using both functional correctness and quantum semantic equivalence. Functional correctness is measured by Pass@K, which evaluates whether at least one generated solution passes the correctness tests. Quantum semantic equivalence is assessed using Process Fidelity, which measures the similarity between the quantum operations implemented by the generated code and the canonical solution.
%
%
We evaluate nine recent LLMs, including general-purpose instruction-tuned models, such as GPT-4.1 \cite{achiam2023gpt}, Claude 3.7 \cite{anthropic_claude}, and Gemini 2.5 \cite{deepmind_gemini}, and code-specialized models, such as CodeLlama \cite{touvron2023llama} and DeepSeek \cite{guo2024deepseek}.

Our results show that, while some models can occasionally generate correct quantum code, their overall capabilities remain limited. The Pass@1 accuracy remains below 40\%, and the highest Pass@5 score reaches only 50\%. Most models solve only a subset of tasks, and several problems remain unsolved by any model. DeepSeek R1 achieves the highest task coverage, particularly in quantum state preparation; however, even the best model reaches only an average Process Fidelity of 51.8\%. Furthermore, model performance does not correlate strictly with model size, indicating that domain-specific adaptation has a stronger impact than model scale alone.

This paper makes the following contributions:
\begin{itemize}
    \item We introduce \textit{QuanBench}, a benchmark for systematically evaluating LLM-based quantum code generation for various quantum programming tasks.
    \item We conduct comprehensive experiments on nine \black{LLMs}, evaluating both functional correctness and quantum semantic equivalence.
    \item We analyze common failure patterns in LLM-generated quantum code, identify key limitations in current models, and provide insights for future improvement.
\end{itemize}

The remainder of this paper is organized as follows. Section~\ref{sec:background} provides background on quantum computing and LLM-based code generation. Section~\ref{sec:constructl} describes the construction of QuanBench. Section~\ref{sec:evaluation} presents the experimental setup, research questions, and evaluation metrics. Section~\ref{sec:experiemntresult} presents the experimental results. Section~\ref{sec: threats} outlines the threats to validity. The related work is discussed in Section~\ref{sec:relatedwork}, and the paper is concluded in Section~\ref{sec:conclusion}.

\section{Background}\label{sec:background}

This section presents the background necessary to support the design and evaluation of QuanBench. We briefly review \deleted{large language models (LLMs)} \black{LLMs}, existing benchmarks for code generation, and the fundamentals of quantum computing relevant to quantum code generation tasks.

\subsection{Large Language Models}

Large language models (LLMs)~\cite{zhao2023survey}, built on the Transformer architecture, have demonstrated strong generalization across various natural language processing tasks, including text summarization, logical reasoning, sentiment analysis, and mathematical problem-solving. Models such as ChatGPT~\cite{ouyang2022training}, DeepSeek~\cite{deepseekai2024deepseekv3technicalreport}, and Gemini~\cite{team2023gemini} are pre-trained on large-scale text corpora and fine-tuned with instruction-following objectives, enabling broad applicability across both general and specialized domains.

Despite these advances, general-purpose LLMs still face limitations in code generation tasks such as program synthesis, bug fixing, and code summarization. For example, LLaMA 2-7B~\cite{touvron2023llama} achieves only 44.4\% Pass@100 on the HumanEval benchmark~\cite{chen2021evaluating}, indicating the challenges in reliably generating functionally correct code.

To address this, code-specific LLMs have been developed~\cite{guo2024deepseek, team2024codegemma}. Models such as StarCoder~\cite{li2023starcoder}, CodeLlama~\cite{roziere2023code}, and Codex~\cite{chen2021evaluating} use domain-specific pre-training and fine-tuning to enhance programming capabilities. For example, CodeLlama-7B achieves 85.9\% Pass@100 on HumanEval~\cite{chen2021evaluating}, showing substantial improvements over general-purpose models. These results highlight the importance of domain specialization in LLM training for code generation tasks.

\subsection{Benchmarks for Code Generation}

Multiple benchmarks have been introduced to evaluate LLMs in code generation. HumanEval~\cite{chen2021evaluating} is widely used, offering short programming problems with test cases to assess functional correctness. CodeContests~\cite{li2022competition} extends evaluation to competitive programming tasks, while EvalPlus~\cite{liu2023your} expands HumanEval with additional test coverage. RMCBench~\cite{chen2024rmcbench} focuses on evaluating LLM resistance against malicious code.

For the quantum domain, Qiskit HumanEval~\cite{vishwakarma2024qiskit} adapts the HumanEval format to quantum programming using Qiskit syntax and constructs. However, it primarily evaluates API usage and syntax compliance rather than algorithmic reasoning or quantum semantic correctness.

To address this gap, we propose QuanBench, a benchmark designed to evaluate the ability of LLMs to generate functionally correct and semantically meaningful quantum code. In contrast to Qiskit HumanEval, QuanBench focuses on algorithmic reasoning and quantum-semantic fidelity, covering a broader set of quantum tasks grounded in real algorithmic challenges.

\subsection{Quantum Computing}

Quantum computing uses principles of quantum mechanics, including superposition, entanglement, and unitary evolution, to process information in a manner distinct from classical computing~\cite{nielsen2010quantum}. A quantum bit (qubit) can exist in a superposition of basis states $|0\rangle$ and $|1\rangle$, represented as:
\begin{equation}
|\psi\rangle = \alpha|0\rangle + \beta|1\rangle,\ \text{where}\ |\alpha|^2 + |\beta|^2 = 1.
\label{equation:quantum_state}
\end{equation}

Quantum computation applies unitary gates to manipulate the qubit states. Gates include single-qubit operations (e.g., Hadamard, Pauli gates) and multi-qubit gates (e.g., CNOT, Toffoli gates), forming quantum circuits that implement algorithms. For an $N$-qubit system, a quantum gate is represented by a $2^N \times 2^N$ unitary matrix.

Several frameworks support the development of quantum programs, including Qiskit~\cite{qiskit2024}, Cirq~\cite{cirq}, and Q\#~\cite{svore2018q}. These toolkits provide abstractions for circuit design, simulation, and execution on both simulators and real quantum hardware.

Classical quantum algorithms such as Grover’s search~\cite{grover1996fast}, Quantum Fourier Transform (QFT)~\cite{weinstein2001implementation}, and Deutsch-Jozsa~\cite{gulde2003implementation} demonstrate computational advantages over classical methods. More recently, parameterized quantum circuits (also known as variational circuits)~\cite{cerezo2021variational} have attracted attention for hybrid quantum-classical computation in applications such as quantum chemistry and optimization.

These properties present unique challenges for code generation models, including handling circuit structure, maintaining quantum state coherence, and correctly composing quantum operators.

Despite these developments, quantum programming remains a challenging task~\cite {ying2016foundations}. Unlike classical programming, the semantics of quantum gate operations depend on the algorithmic context, making program design, reasoning, and automation complex.

In parallel, LLMs have shown strong capabilities in classical code generation, raising the question of whether LLMs can also handle quantum program semantics and algorithmic intent. To investigate this, we introduce QuanBench, a benchmark designed to systematically assess LLMs’ ability to generate quantum programs across diverse algorithmic tasks, using semantic evaluation metrics and canonical solutions, to assess both algorithmic reasoning and quantum semantic correctness.

\section{Benchmark Construction}\label{sec:constructl}

\begin{figure*}
    \centering
    \includegraphics[width=0.95\linewidth]{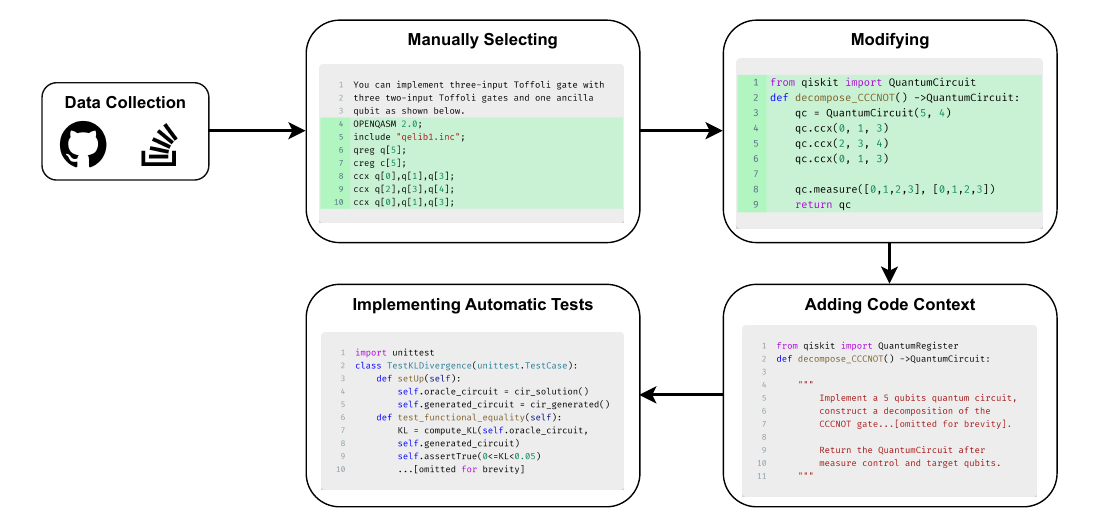}
    \caption{QuanBench Design}
    \label{fig:design}
\end{figure*}

The construction of QuanBench consists of three stages, as shown in Figure~\ref{fig:design}. 

\begin{itemize}
\item First, we curated a diverse set of high-quality quantum algorithm implementations by collecting programs from open-source GitHub repositories and extracting relevant issues from StackOverflow. 

\item Second, each collected item was reformulated as a clearly defined programming problem with a corresponding canonical solution. All canonical solutions are executable and faithfully represent the intended quantum algorithms. 

\item Third, we implemented a multi-criteria evaluation pipeline for each benchmark problem, which includes functional correctness checks, comparative testing, and constraint-based validation. This evaluation framework allows us to assess both the syntactic correctness and the quantum semantic equivalence of the model-generated code.
\end{itemize}

\subsection{Problem Selection}

\begin{figure*}
    \centering
    \includegraphics[width=1.02\linewidth]{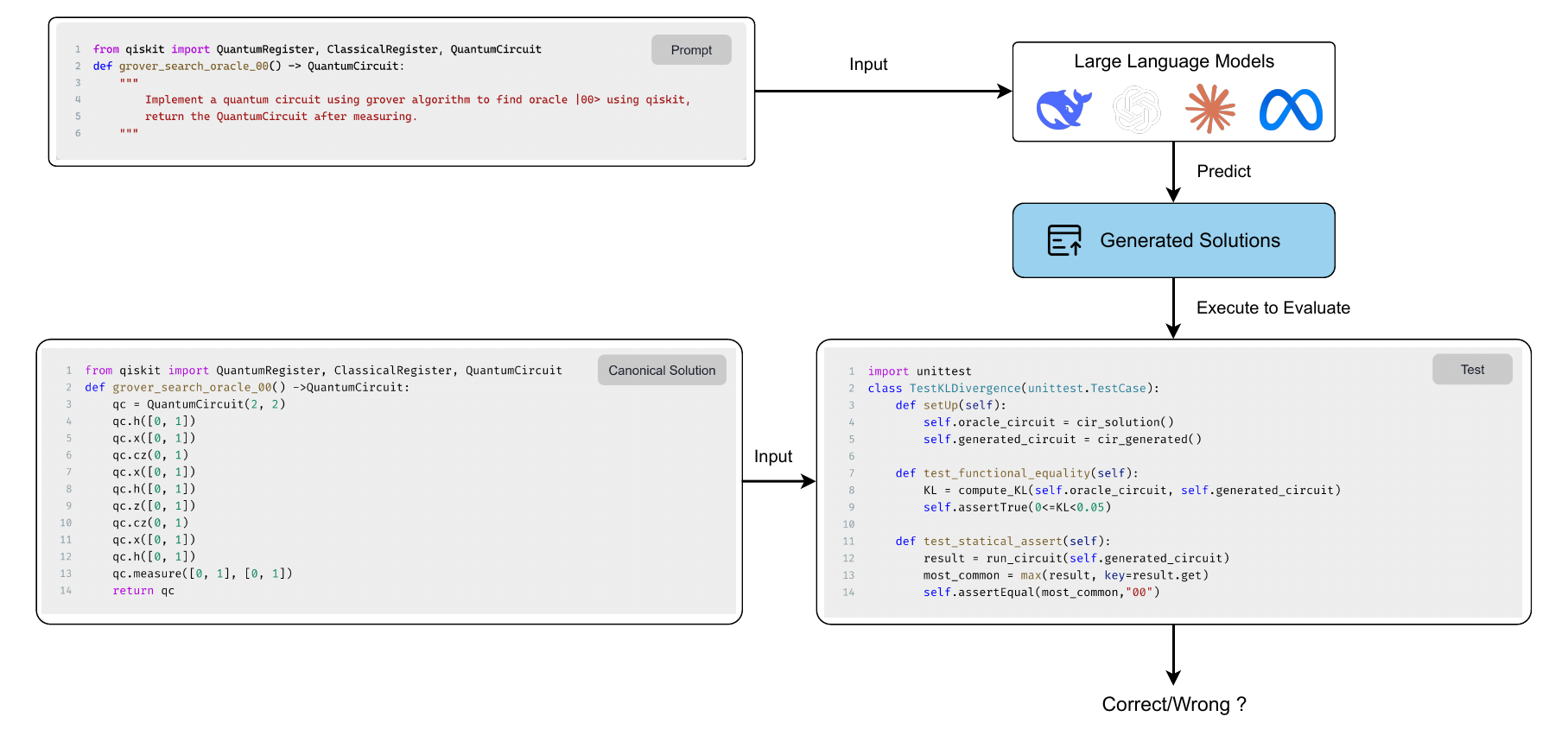}
    \caption{One Example in QuanBench}
    \label{fig:example}
\end{figure*}

\black{\textbf{$\bullet$ Sourcing Problems.} To collect high-quality benchmark problems, we retrieve repositories containing quantum program content by searching for keywords "Quantum algorithm" and "Qiskit" from GitHub (questions from Stack Overflow), using the following filtering process:}

\begin{itemize}
    \item \black{First, temporal filtering was applied by including only repositories created after 2023, which helps reduce potential overlap with the training data of current \deleted{large language models (LLMs)} \black{LLMs}, whose datasets primarily cover pre-2023 content.}
    \item \black{Second, popularity-based filtering is applied: GitHub repositories with more than 5 stars and Stack Overflow posts with more than 50 views are retained. Using this threshold as a preliminary proxy for community validation regarding utility and correctness.}
\end{itemize}

\black{This process collected a total of 127 GitHub repositories and 39 Stack Overflow questions, which were then manually screened for quality and diversity.}

\black{\textbf{$\bullet$ Filtering Suitable Problems.} From the initial repository and program pool, we applied additional manual screening based on the following criteria:}

\begin{itemize}
    \item \black{To eliminate ambiguity, each program must be accompanied by a clear and well-defined problem description, and testable behaviors.}
    \item \black{Each collected program is required to successfully execute without errors and to yield correct outputs in accordance with its behavior.}
    \item \black{To eliminate redundancy, programs that employed the same algorithmic approach to address the same problem were excluded.}
\end{itemize}

\black{Furthermore, we prefer programs that demonstrate scalability and extensibility, meaning that they can be generalized to handle larger numbers of qubits and be adapted to a broader range of quantum tasks. After applying these criteria, we curated a final set of 44 unique quantum algorithm programs to include in the benchmark suite. (42 from GitHub and 2 from Stack Overflow)}

\subsection{Rewriting Problems and Canonical Solutions}

\noindent
\black{\textbf{$\bullet$ Version Control.}
Since the training data of LLMs such as ChatGPT typically extends only up to 2023, these models often lack familiarity with subsequent versions of Qiskit. To ensure environmental consistency and minimize potential confounding effects on LLM performance, we standardized the environment to Qiskit 0.46.0 and Python 3.10. This choice does not materially affect either the prompts or the solutions, as our tasks rely solely on the core quantum circuit logic, which remains compatible across different versions. The only differences arise in the construction and execution of the evaluation metrics (e.g., simulators and transpilation tools), which rely on consistent backend behavior.}

\noindent
\black{\textbf{$\bullet$ Translation.}
Since not all programs were originally implemented in Qiskit, some were written in frameworks such as Cirq or Classiq, and we manually translated them into Qiskit. To ensure the correctness of the translation, we manually verify that the structure of the translated quantum circuit matches the original and that both produce identical outputs.
}

\noindent
\black{\textbf{$\bullet$ Core Circuit Logic Extraction.}
QuanBench is designed around Qiskit-based quantum programming tasks, which may include simulation or hybrid optimization logic beyond the quantum circuit itself (e.g., QAOA integrates classical optimization). Due to substantial syntax differences across Qiskit versions, we manually extract only the circuit part from the translated programs. To ensure clarity and consistency, simulation and hybrid optimization logic are implemented within the test code, isolating them from the main quantum circuit. The docstring specifies that the output should be a QuantumCircuit.
}

\noindent
\black{\textbf{$\bullet$ Canonical Solutions}
The extracted circuits serve as our canonical solutions. Each is validated using test cases to ensure functional equivalence with the original program.
}

\subsection{\black{Docstring Construction}}
\black{We wrote docstrings based on the extracted core circuit logic, covering the problem description, qubit allocation, algorithm used, and measurement strategy. All docstrings were cross-validated by three experts in quantum programming. Subsequently, we used ChatGPT to generate a program from the docstring to confirm clarity and reduce the chance of misinterpretation.}

\subsection{Implementing Multi-Criteria Evaluations}

To evaluate the correctness of quantum programs generated by \deleted{large language models (LLMs)} \black{LLMs}, we adopt a multi-criteria evaluation framework that combines probabilistic output comparison and static constraint validation. This framework captures both the functional and structural correctness of the generated code.

\vspace*{1mm}
\noindent
\textbf{$\bullet$ KL Divergence.}  
For tasks involving measurement outcomes (e.g., quantum teleportation \cite{bouwmeester1997experimental} and Bell state preparation), we compare the output probability distributions of the generated code and the canonical solution using the Kullback-Leibler (KL) divergence. This metric quantifies the divergence between the model's output distribution $P$ and the reference distribution $Q$:
\begin{equation}
        D_{\text{KL}}(P \parallel Q) = \sum_{x} p(x) \log \frac{p(x)}{q(x)}.
\end{equation}
A lower KL divergence indicates that the model-generated program produces output distributions statistically closer to the correct quantum implementation.

As shown in Figure~\ref{fig:example}, we use Python’s \texttt{unittest} framework to verify functional correctness. Specifically, we apply the \texttt{assertTrue} statement with a threshold of 0.05: the assertion passes only if the KL divergence between the generated and reference distributions is below 0.05.

\vspace*{1mm}
\noindent
\textbf{$\bullet$ Static Constraint Validation.}  
In addition to output-based evaluation, quantum programs must satisfy certain structural and semantic constraints to be considered valid. In QuanBench, we define multiple static constraints, including:

\begin{itemize}
    \item \textbf{Gate set constraints}: The set of gates used in the generated circuit must be a subset of those used in the reference implementation.
    \item \textbf{Measurement result constraints}: The output distribution must include specific target states and reflect expected probability relationships.
    \item \textbf{Phase constraints}: The generated circuit must preserve relative phases of quantum states where applicable.
\end{itemize}

Figure~\ref{fig:example} illustrates an example of a measurement result constraint. In this case, for a Grover algorithm marking the state $|00\rangle$, the correct behavior is reflected in the output distribution where state $00$ should have the highest probability. The generated circuit is accepted only if it satisfies this expected pattern.

\subsection{Dataset Statistics}
\label{subsec: require}
\deleted{The current version of QuanBench comprises 44 curated tasks that cover a diverse range of quantum programming problems. Table~\ref{tab: cat} summarizes the task distribution across four categories, based on their quantum computation objectives. This design aims to balance both high-level algorithmic reasoning and low-level circuit construction skills required in quantum programming.}

\black{QuanBench was developed to assess LLM's ability to handle often occurring quantum programming tasks based on natural language task descriptions. Table~\ref{tab: cat} summarizes the task distribution in four categories, defined according to their underlying quantum computational objectives. In constructing QuanBench, we aimed to include as many commonly used and practically relevant quantum algorithms as possible (e.g., HHL, QFT, Grover's, and Shor's algorithms) and balance both high-level algorithmic reasoning and low-level circuit construction skills required in quantum programming.}

\black{Scalability was another key consideration: tasks were selected for their potential to be extended to more qubits or restructured to accommodate alternative problem formulations. Following these design criteria, we assembled 44 quantum programming tasks for the QuanBench benchmark.}

\vspace*{1mm}
\noindent
\textbf{$\bullet$ Quantum Algorithm.}  
This category encompasses full implementations of established quantum algorithms that demonstrate quantum computational advantages, including Grover's search~\cite{grover1996fast}, Shor's factoring~\cite{monz2016realization}, Deutsch-Jozsa~\cite{gulde2003implementation}, Bernstein-Vazirani~\cite{arvind2007optical}, and Simon's algorithms~\cite{brassard1997exact}. Tasks in this category involve multiple components, including state initialization, oracle construction, and algorithm-specific logic. These problems evaluate whether LLMs can understand the overall structure and procedural flow of complete quantum algorithms.

\vspace*{1mm}
\noindent
\textbf{$\bullet$ Quantum State Preparation.}  
This category focuses on constructing specific quantum states and analyzing their structural or entanglement properties. Tasks include preparing states such as Bell, GHZ, and W states, which require accurate encoding of entanglement and superposition. These problems assess whether LLMs can translate target states into correct gate-level implementations.

\vspace*{1mm}
\noindent
\textbf{$\bullet$ Gate Decomposition.}  
These tasks require decomposing high-level quantum operations into sequences of elementary gates \cite{vartiainen2004efficient}, which are supported by the hardware. Gate decomposition is crucial for the practical deployment of current quantum devices. The tasks assess the understanding of the quantum gate algebra, the decomposition rules (for example, the Toffoli gate decomposition), and the ability to maintain the correctness of the circuit during transformation.

\vspace*{1mm}
\noindent
\textbf{$\bullet$ Quantum Machine Learning.}  
This emerging category includes tasks involving parameterized quantum circuits (PQCs) \cite{benedetti2019parameterized}, which are widely used in hybrid quantum-classical learning algorithms. PQCs are applied in tasks such as classification, regression, and quantum state discrimination. These problems test whether LLMs can synthesize circuit templates that correctly implement variational models, even when the problem specifications are incomplete or abstract.

This category design enables QuanBench to address a broad spectrum of quantum programming challenges, ranging from classical algorithmic reasoning to low-level circuit manipulation and the synthesis of quantum components on hybrid quantum-classical models. The combination provides a comprehensive evaluation setting for analyzing the quantum programming capabilities of LLMs.

{\subsection{\black{Dataset Expansion}}
\black{QuanBench currently includes the majority of common quantum algorithms and supports scalable task design, but full balance across task types has not yet been achieved. The benchmark is actively expanding to include variations of existing algorithms across different qubit sizes, adaptations to new problem types, and tasks related to Quantum Machine Learning. Table~\ref{tab: cat} shows that the current stage of QuanBench comprises 117 tasks, with improved task diversity and relative category balance.}

\black{The current version of QuanBench is scheduled for open source release in the near future, and the benchmark remains under active development and expansion}


\begin{table}[h]
\centering
\caption{Task Categorization in QuanBench}
\label{tab: cat}
\begin{tabular}{lcc}
\hline
\multirow{2}{*}{\textbf{Category}} & \multicolumn{2}{c}{\textbf{Number of Tasks}}                      \\ \cline{2-3} 
                                   & \multicolumn{1}{l}{QuanBench} & \multicolumn{1}{l}{Current Stage} \\ \hline
Quantum Algorithm                  & 32                            & 64                                \\
Quantum State Preparation          & 6                             & 24                                \\
Gate Decomposition                 & 5                             & 14                                \\
Quantum Machine Learning           & 1                             & 15                                \\ \hline
Total                              & 44                            & 117                               \\ \hline
\end{tabular}
\end{table}

\section{Evaluation of LLM Models}\label{sec:evaluation}

\subsection{Research Questions (RQs)}

We evaluate the capabilities of LLMs on quantum algorithm generation tasks through the following research questions:

\begin{itemize}
    \item \textbf{RQ1:} What is the current capability of \deleted{large language models (LLMs)}\black{LLMs} in generating quantum algorithm programs? This question focuses on the overall performance of LLMs in producing correct, functional, and executable quantum code from natural language descriptions.
    
    \item \textbf{RQ2:} What is the gap between LLM-generated quantum algorithms and canonical solutions? This question examines the semantic and structural differences between model-generated code and reference implementations.
    
    \item \textbf{RQ3:} What are the common causes of errors in LLM-generated quantum programs? This question analyzes both qualitative and quantitative error patterns, identifying failure modes such as deprecated APIs, incorrect qubit indexing, and semantic deviations.

    \item \textbf{\black{RQ4:}} \black{What is the effect of temperature and batch size on the performance of LLMs in QuanBench? The question aims to identify suitable parameter settings for fair and efficient benchmarking and to provide insights that can guide both future research and the practical optimization of LLMs for quantum code generation.}

\end{itemize}

\subsection{Model Selection and Settings}

\vspace*{1mm}
\noindent
\textbf{$\bullet$ Model Selection.} To evaluate the quantum code generation capabilities of LLMs, we selected a diverse set of models that have demonstrated strong performance in code generation tasks. The selected models are listed in Table~\ref{tab:sel-models}.


\begin{table}[ht]
\caption{List of Evaluated LLMs in QuanBench}
\label{tab:sel-models}
\centering
\begin{tabularx}{\linewidth}{l X l c c}
\toprule
\textbf{Category} & \textbf{LLM} & \textbf{Organization} & \textbf{Year} & \textbf{Open} \\
\midrule
\multirow{7}{*}{Gen. LLM} 
  & GPT-4.1 \cite{achiam2023gpt} & OpenAI & 2025 & no \\
  & DeepSeek V3 \cite{deepseekai2024deepseekv3technicalreport} & DeepSeek-AI & 2024 & yes \\
  & DeepSeek R1 \cite{guo2025deepseek} & DeepSeek-AI & 2025 & yes \\
  & Claude 3.7 Sonnet \cite{anthropic_claude} & Anthropic & 2025 & no \\
  & Gemini 2.5 \cite{deepmind_gemini} & Google & 2025 & no \\
  & Llama-4-Maverick-17B \cite{touvron2023llama} & Meta & 2025 & yes \\
  & Qwen2.5-7B-Instruct \cite{hui2024qwen2} & Qwen & 2025 & yes \\
\midrule
\multirow{2}{*}{Code LLM} 
  & CodeLlama-7B \cite{roziere2023code} & Meta & 2023 & yes \\
  & CodeLlama-34B \cite{roziere2023code} & Meta & 2023 & yes \\
\bottomrule
\end{tabularx}
\end{table}

The model selection followed the criteria below:

\begin{itemize}
    \item We prioritized models with demonstrated strong performance in code generation, particularly those fine-tuned on programming or instruction datasets.
    \item We excluded open-source models that lack version identifiers or checkpoint information to ensure full reproducibility.
    \item \black{We include models with more than 7B parameters only if their inference APIs are available, due to local hardware inference limitations.}
    \item Following prior studies~\cite{vishwakarma2024qiskit}, which showed that small-scale LLMs perform poorly in quantum programming tasks, we focused on medium-to-large models, including 30B+ instruction-tuned variants.
\end{itemize}

This selection strategy ensures a well-balanced model set that allows meaningful comparison across model size, architecture, and deployment modes.

\vspace*{1mm}
\noindent
\textbf{$\bullet$ Model Settings.}
To ensure fair and reproducible evaluation, we applied unified inference configurations across all models and maintained controlled execution environments. Proprietary models (e.g., GPT-4.1, Claude 3.7 Sonnet) were accessed via official APIs, while open-source models were executed locally or via Hugging Face Inference Endpoints.

For open-source models such as CodeLLaMA-7B and Qwen2.5-7B-Instruct, we downloaded official model weights from Hugging Face repositories and ran inference locally. The proprietary models were accessed through their official providers or authorized third-party APIs, including OpenAI, DeepSeek, Anthropic, Google, Novita, and Replicate.

All models were evaluated using consistent inference parameters, with the temperature set to 0.8. We used a HumanEval-style prompt format to standardize task presentation across models. 

\subsection{Evaluation Metric}

To comprehensively assess both the correctness and the quantum functionality of the code generated by \deleted{large language models (LLMs)}\black{LLMs}, we adopt two complementary evaluation metrics. Pass@K and Process Fidelity. These metrics jointly capture syntactic/functional correctness as well as quantum semantic equivalence.

Similar to HumanEval~\cite{chen2021evaluating}, we use Pass@K to measure the probability that an LLM generates at least one correct solution among its top-K outputs for a given problem. Specifically, for each of the 44 benchmark problems in \textit{QuanBench}, Pass@K reports the proportion of problems where at least one of the K generated samples \deleted{passes}\black{pass} all correctness tests, including static verification and simulation-based evaluation.

In our experiments, each model generates $n=5$ samples per problem. We report both Pass@1 and Pass@5 to assess model performance under single-sample and multi-sample settings. To reduce sampling variance, we apply the unbiased estimator of Pass@K used in HumanEval~\cite{chen2021evaluating}, defined as:
\begin{equation}
\text{Pass@}K = 1 - \frac{\binom{n - C}{K}}{\binom{n}{K}},
\end{equation}
where $C$ is the number of correct generations. This formulation enables statistically robust comparisons across models with varying generation quality.

In addition to Pass@K, we introduce Process Fidelity to evaluate the quantum semantic correctness of generated circuits. Unlike Pass@K, which is based on test outcomes, Process Fidelity directly measures the similarity between the quantum operations implemented by the generated circuit and the canonical solution.

Formally, for each benchmark problem, let $U_\text{ref}$ be the unitary matrix of the canonical solution, and $U_\text{gen}$ be the unitary matrix of the generated circuit. The process fidelity $\mathcal{F}$ is calculated as:
\begin{equation}
\mathcal{F}(U_\text{ref}, U_\text{gen}) = \left|\frac{1}{d} \mathrm{Tr}\left(U_\text{ref}^\dagger U_\text{gen}\right)\right|^2,
\end{equation}
where $d = 2^n$ is the Hilbert space dimension for a $n$-qubit system, and $\mathrm{Tr}(\cdot)$ denotes the matrix trace. A fidelity score of 1.0 indicates exact functional equivalence up to a global phase.

In our experiments, for each of the 44 benchmark problems, fidelity is computed between the canonical circuit and each of the 5 generated circuits, yielding 220 fidelity comparisons per model. This metric enables fine-grained evaluation of whether the generated circuit preserves the intended quantum transformation, regardless of low-level variations such as gate decomposition or gate ordering.

Together, Pass@K and Process Fidelity provide a robust framework to assess both the functional correctness and the quantum mechanical soundness of LLM-generated quantum programs.

\section{Experimental Results}\label{sec:experiemntresult}
In this section, we present the experimental results of the evaluation of selected LLMs on \textit{QuanBench}. The analysis addresses each research question (RQ1–RQ4) and provides quantitative and qualitative insights into the quantum code generation capabilities of the models.

\subsection{RQ1: What is the current capability of \deleted{large language models (LLMs)} \black{LLMs} in generating quantum algorithm programs?}

\subsubsection{Overall Performance}

Figure~\ref{fig:passk} presents the performance of the LLMs evaluated in QuanBench. The x-axis lists the models, and the y-axis shows their Pass@K scores. Both Pass@1 and Pass@5 results are reported for each model.

Across all models, the highest Pass@1 accuracy does not exceed 0.4, and the best Pass@5 accuracy reaches only 0.5, indicating that current LLMs have limited capability to generate correct quantum algorithm implementations. Among the models, DeepSeek R1 and Gemini 2.5 achieve the highest performance, with Pass@1 = 0.38 and Pass@5 = 0.5, respectively.

In contrast, CodeLlama-7B shows the weakest performance, with a Pass@5 of only 0.11. Models such as GPT-4.1, DeepSeek V3, and Claude 3.7 produce comparable results. In particular, CodeLlama-34B performs between Qwen2.5-7B and Llama-4-Maverick-17B, suggesting that the performance of quantum code generation does not strictly correlate with the size of the model, and CodeLlama models generally underperform on these tasks.

\begin{figure*}
    \centering
    \includegraphics[width=0.75\linewidth]{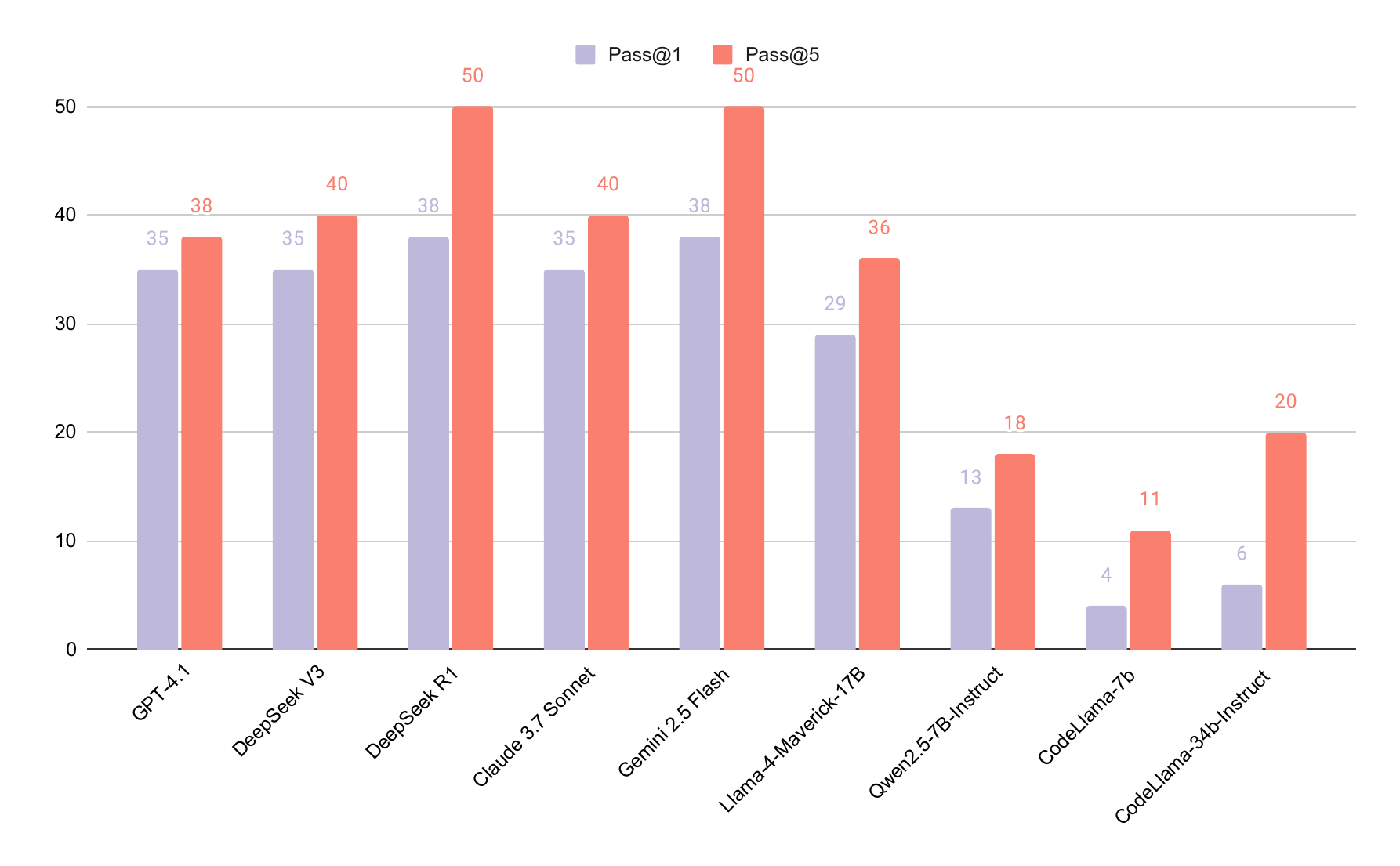}
    \caption{Effectiveness comparison between LLMs on QuanBench}
    \label{fig:passk}
\end{figure*}

\subsubsection{Task Coverage}

We further analyzed task coverage by counting the number of problems for which each model generated at least one correct solution within its five generated samples. This provides a complementary perspective to Pass@K by focusing on success at the task level rather than per-sample accuracy.

As shown in Figure~\ref{fig:solvedproblems}, the scope of tasks varies significantly between models. DeepSeek R1 and Gemini 2.5 achieve the highest coverage, solving 22 out of 44 problems each. Notably, DeepSeek R1 uniquely solves three tasks that no other model completed, and it is the only model that successfully solves all Quantum State Preparation tasks, indicating particular strength in that category.

In contrast, other models perform well only on narrower subsets of tasks, with generally lower success rates for state preparation and gate decomposition problems. These results indicate that performance differences among LLMs extend beyond aggregate Pass@K scores and reflect variations in their ability to handle diverse quantum programming tasks.

\begin{figure*}
    \centering
    \includegraphics[width=0.8\linewidth]{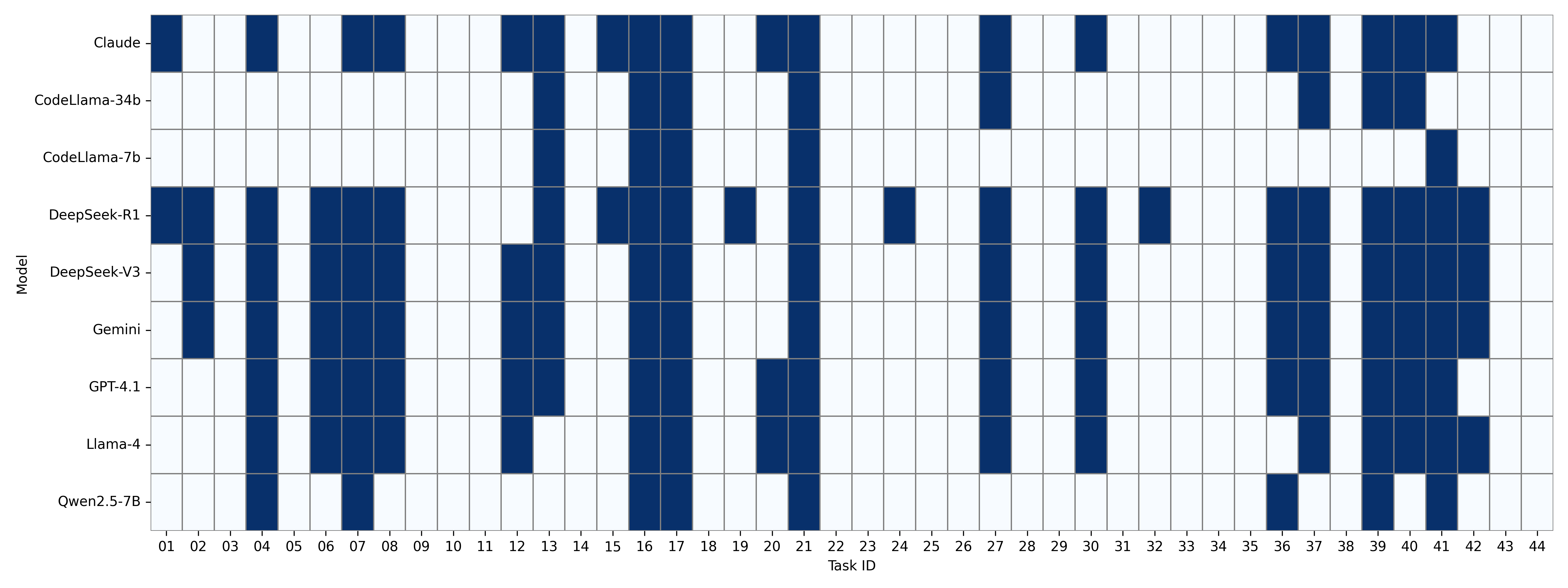}
    \caption{Number of problems solved on QuanBench}
    \label{fig:solvedproblems}
\end{figure*}

\begin{tcolorbox}[size=title,rightrule=1mm, leftrule=1mm, toprule=0mm, bottomrule=0mm, arc=0pt,colback=gray!5,colframe=bleudefrance!75!black,breakable]
\textbf{Answer to RQ1:} Overall, current LLMs exhibit limited capability to generate complete and correct implementations of quantum algorithms. Although some models show strength in specific categories, their performance does not consistently generalize across the diverse set of quantum programming tasks in QuanBench. Different models display different strength patterns, suggesting varying internal representations and reasoning strategies for quantum circuits.
\end{tcolorbox}

\subsection{RQ2: What is the gap between LLM-generated quantum algorithms and canonical solutions?}

To evaluate the fidelity and structural correctness of quantum circuits generated by \deleted{large language models (LLMs)}\black{LLMs}, we introduced the Process Fidelity metric, which quantifies the similarity between a generated circuit and its canonical counterpart in terms of their underlying quantum transformations. During evaluation, generated circuits that failed to compile were excluded.

As shown in Table~\ref{tab:process_fidelity}, the average process fidelity scores across models are significantly lower than 100\%, indicating that most generated circuits diverge from the intended quantum operations. While larger models tend to achieve higher fidelity (typically around 50\%), GPT-4.1 achieves only 47\%, consistent with its relatively low Pass@K performance.

\begin{table}[h!]
\centering
\caption{Process Fidelity Scores of Different LLMs}
\label{tab:process_fidelity}
\begin{tabular}{@{}lccc@{}}
\toprule
\textbf{LLM} & \textbf{Passed} & \textbf{Failed} & \textbf{Average} \\
\midrule
GPT-4.1 & 70 & 28 & 47 \\
DeepSeek V3 & 75 & 33 & 51 \\
DeepSeek R1 & 49 & 51 & 50 \\
Claude 3.7 Sonnet & 75 & 32 & 52 \\
Gemini 2.5 Flash & 57 & 41 & 50 \\
Llama-4-Maverick-17B & 60 & 30 & 41 \\
Qwen2.5-7B-Instruct & 64 & 8 & 20 \\
CodeLlama-7b & 65 & 5 & 14 \\
CodeLlama-34b-Instruct & 72 & 10 & 24 \\
\bottomrule
\end{tabular}
\end{table}

To gain deeper insight into the relationship between functional correctness and quantum-semantic similarity, we separately computed Process Fidelity for generated circuits that pass and fail the test cases.

Our results show that for models such as GPT-4.1, DeepSeek V3, and Claude 3.7, the \deleted{average process fidelity among the test-passing samples}\black{passed process fidelity} reaches approximately 70\%, indicating that the generated circuits are structurally and semantically close to the canonical solutions. In contrast, their failing samples exhibit much lower fidelity (around 30\%), suggesting a significant divergence from the intended quantum behavior.

Interestingly, for larger models, such as DeepSeek R1 and Gemini 2.5, the process fidelity remains consistently around 50\% for both successful and failed samples. Further inspection of some low-fidelity but functionally correct samples reveals that discrepancies often arise from structural variations that do not affect the final measurement results. These include differences in qubit ordering and global or relative phase shifts that lead to identical observable behavior.

For example, as shown in Figure~\ref{fig:GHZ}, the construction of a GHZ state can be implemented using different but functionally equivalent circuit structures. One approach applies CNOT gates from the initial qubit in superposition to all other qubits in parallel, while another applies CNOT gates sequentially from one qubit to the next in a chain-like manner. Although both implementations generate the same entangled state, their circuit representations differ, which may result in lower process fidelity despite being functionally correct.

\begin{figure}
    \centering
    \hspace{-0.5cm}
    \includegraphics[width=1.05\linewidth]{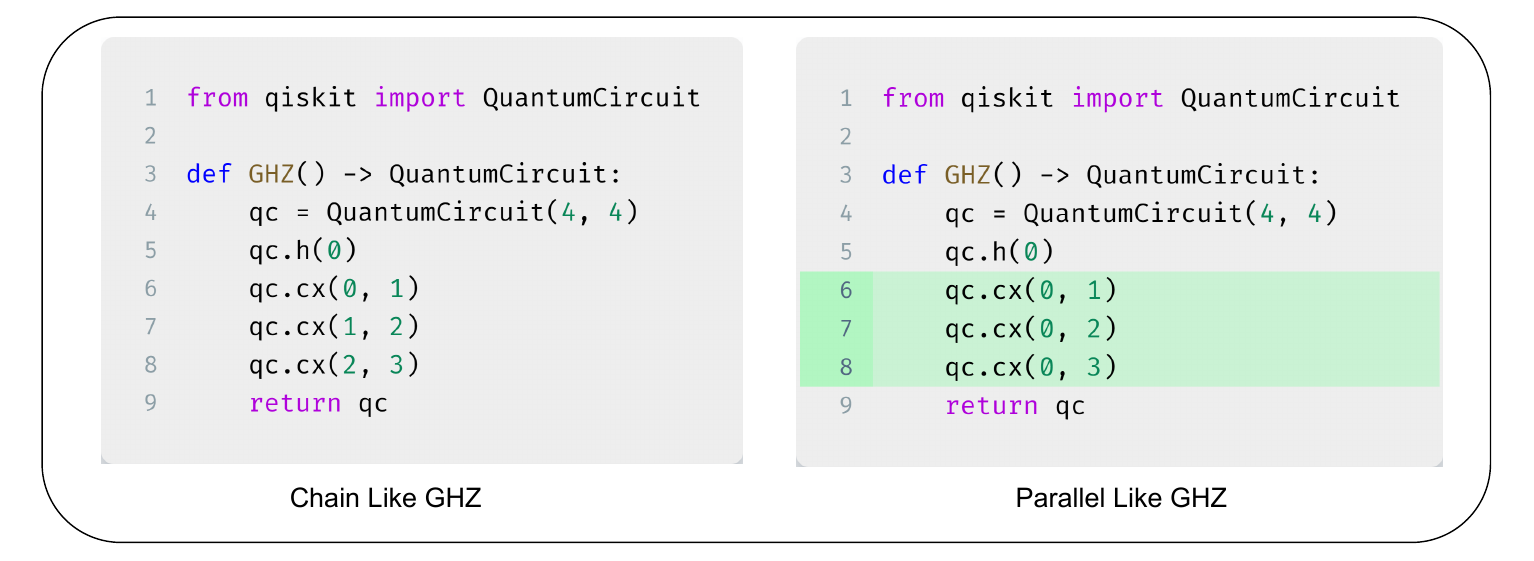}
    \caption{Different Constructions of GHZ State}
    \label{fig:GHZ}
\end{figure}

We hypothesize that this behavior reflects the partial ability of these models to capture deeper quantum algorithmic structures. As a result, even some incorrect samples may exhibit unitary behavior closer to the reference circuit, while successful samples may achieve correctness through alternative circuit constructions, resulting in lower process fidelity despite functional equivalence. This suggests that stronger models may generalize better by producing semantically diverse but functionally valid circuits, an important trait for practical quantum code generation.

\begin{tcolorbox}[size=title,rightrule=1mm, leftrule=1mm, toprule=0mm, bottomrule=0mm, arc=0pt,colback=gray!5,colframe=bleudefrance!75!black,breakable]
\textbf{Answer to RQ2:} This highlights a critical limitation: current LLMs often succeed in capturing surface-level functional objectives, but struggle to preserve deeper quantum semantics, such as correct operator composition, entanglement structures, or unitary decompositions. Bridging this semantic gap is essential to achieve reliable and generalizable quantum code generation.
\end{tcolorbox}

\subsection{RQ3: What are the common causes of errors in LLM-generated quantum programs?}

To gain insight into the failure modes of LLM-generated quantum circuits, we conducted a detailed error analysis of the failed samples. Several recurring patterns were identified that significantly contribute to the observed failure rates.

\vspace*{1mm}
\noindent
\textbf{$\bullet$ Use of Deprecated Gate APIs.} A notable portion of compilation failures stemmed from the use of outdated gate APIs. For example, the \texttt{cu1} gate has been deprecated in Qiskit versions after 0.45.0, replaced by \texttt{cp}, or alternatively accessible via \texttt{qiskit.circuit.library.CU1Gate}. All evaluated models, except Gemini 2.5, DeepSeek V3, and DeepSeek R1, incorrectly invoked the deprecated \texttt{cu1} gate, resulting in compilation errors. This indicates that several LLMs rely on outdated training data and do not know about version-specific API updates.

\vspace*{1mm}
\noindent
\textbf{$\bullet$ Structural Errors in Circuit Construction.}  
Many errors were caused by incorrect qubit indexing, such as assigning the same qubit as both control and target in a controlled operation (e.g., a \texttt{cx} gate with identical indices). This pattern was observed across all models, indicating a limited understanding of quantum gate constraints and circuit construction rules.

\vspace*{1mm}
\noindent
\textbf{$\bullet$ Overuse or Misuse of Gates.}  
In some cases, circuits exhibited high process fidelity, but failed test cases due to extraneous gate operations inserted at the beginning or end of the circuit. Although the core algorithmic structure was correct, redundant gates altered the expected measurement outcomes. This behavior reflects an incomplete grasp of quantum circuit minimality and end-to-end semantics.

\vspace*{1mm}
\noindent
\textbf{$\bullet$ Semantic Deviations in Logic.}  
Even syntactically valid circuits frequently implement incorrect algorithmic logic. For example, a common failure involved improperly constructed oracles in Grover's algorithm, leading to significant drops in process fidelity despite syntactic correctness. These errors suggest that LLMs frequently struggle to comprehend the algorithmic intent and structural requirements of quantum programs.

\begin{figure}
    \centering
    \includegraphics[width=0.85\linewidth]{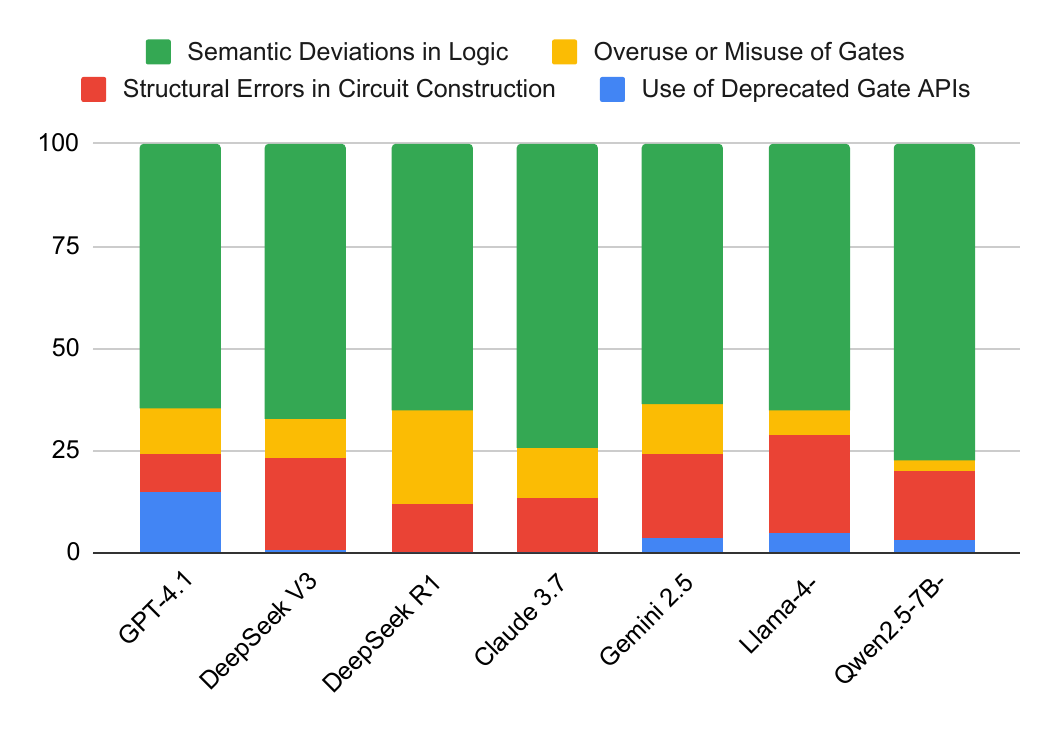}
    \caption{Distribution of identified error types across all models}
    \label{fig:errdis}
\end{figure}

We summarize the distribution of the four identified error types in Figure~\ref{fig:errdis}. Among them, \textbf{Semantic Deviations in Logic} account for the largest proportion of errors, particularly in cases where models fail to generate any correct solution. These errors often involve incorrect oracle construction or missing gate operations, especially in complex tasks such as state preparation and gate decomposition.

The second most frequent error is \textbf{Structural Errors in Circuit Construction}, which often leads to compilation failures due to issues such as invalid qubit indexing or incorrect gate syntax.

\textbf{Overuse or Misuse of Gates} ranks third. In these cases, generated circuits may pass partial tests, but fail under stricter correctness checks. For example, in quantum-state preparation, the absence of necessary phase gates may lead to failures when relative phase relationships are tested. Similarly, in Grover’s algorithm, incorrectly inserted Hadamard gates may destroy the amplitude amplification effect, causing functional errors even when the overall structure appears plausible.

The least frequent error is \textbf{Use of Deprecated Gate APIs}, which is mitigated mainly by standardizing the Qiskit version across evaluations. However, we observe a notable exception with GPT-4.1, which frequently invokes outdated gates such as \texttt{cu1}. We hypothesize that this is due to stale training data or limited exposure to updated quantum programming resources.

\begin{tcolorbox}[size=title,rightrule=1mm, leftrule=1mm, toprule=0mm, bottomrule=0mm, arc=0pt,colback=gray!5,colframe=bleudefrance!75!black,breakable]
\textbf{Answer to RQ3:} Overall, these patterns highlight that while LLMs can generate plausible quantum code snippets, they often lack robust semantic grounding and version-aware reasoning, both of which are critical for producing correct and executable quantum programs. Semantic deviations remain the most critical and challenging issue for LLMs in generating quantum programs.
\end{tcolorbox}

\subsection{\black{RQ4: What is the effect of temperature and batch size on the performance of LLMs in QuanBench}}


\black{Given that LLMs demonstrate different levels of output diversity depending on the temperature setting~\cite{peeperkorn2024temperature}, and that batch size can also affect diversity, we performed experiments to assess their performance on quantum programming tasks under multiple temperature settings and to investigate the influence of batch size when temperature is held constant.}

\vspace*{1mm}
\noindent
\black{\textbf{$\bullet$ Temperature Settings.}}
\black{We first examined the performance of LLMs under different temperature settings, specifically 0.1, 0.5, 0.8, and 1.0. It should be noted that for certain models, the maximum temperature setting is restricted to 1, which restricts the range of diversity experiments that can be conducted. Tables~\ref{tab:temperature-performance1} and~\ref{tab:temperature-performance5} report the results for Pass@1 and Pass@5, respectively, with the best results highlighted in green.}

\black{As shown in Table~\ref{tab:temperature-performance1}, the Pass@1 results indicate that a temperature of 0.8 yields the highest accuracy for most models. For example, DeepSeek V3 achieves its best performance at a temperature of 0.1 with an accuracy of 36\%, which is only marginally higher than its score of 35\% at 0.8. In contrast, CodeLlama models achieve their best results at a temperature of 0.5, which we hypothesize may be related to the model configuration. Overall, for Pass@1, most LLMs perform best when the temperature is set to 0.8.}

\black{In Table~\ref{tab:temperature-performance5}, we again observe that most LLMs achieve their best performance at a temperature of 0.8. At the same time, some models exhibit identical performance across different temperatures (e.g., DeepSeek V3, Gemini, and CodeLlama), while others perform better at a temperature of 1.0. These results suggest that with the current batch size of 5, a temperature of 0.8 generally leads to superior performance for most models.}


\begin{table}[htbp]
\centering
\caption{LLMs Pass@1 performance across different temperatures}
\label{tab:temperature-performance1}
\begin{tabular}{lcccc}
\toprule
\multirow{2}{*}{\textbf{Model}} & \multicolumn{4}{c}{\textbf{Temperature}}                  \\ \cline{2-5} 
                                & \textbf{0.1} & \textbf{0.5} & \textbf{0.8} & \textbf{1.0} \\ \midrule
GPT-4.1                         & 34           & 33           & \cellcolor{green!20}\textbf{35}           & 34           \\
DeepSeek V3                     & \cellcolor{green!20}\textbf{36}           & 35           & 35           & 33           \\
DeepSeek R1                     & 35           & 37           & \cellcolor{green!20}\textbf{38}           & 35           \\
Claude 3.7 Sonnet               & 33           & 33           & \cellcolor{green!20}\textbf{35}           & 32           \\
Gemini 2.5 Flash                & 35           & 36           & \cellcolor{green!20}\textbf{38}           & 36           \\
Qwen2.5-7B-Instruct             & 10           & 10           & \cellcolor{green!20}\textbf{13}           & 8            \\
CodeLlama-7B                    & 4            & \cellcolor{green!20}\textbf{7}            & 4            & 3            \\
CodeLlama-34B-Instruct          & 10           & \cellcolor{green!20}\textbf{11}           & 6            & 6            \\ \bottomrule
\end{tabular}
\end{table}


\begin{table}[htbp]
\centering
\caption{LLMs Pass@5 performance across different temperatures}
\label{tab:temperature-performance5}
\begin{tabular}{lcccc}
\toprule
\multirow{2}{*}{Model} & \multicolumn{4}{c}{Temperature}                                                                                           \\ \cline{2-5} 
                       & \textbf{0.1}                 & \textbf{0.5}                 & \textbf{0.8}                 & \textbf{1.0}                 \\ \midrule
GPT-4.1                & \cellcolor{green!20}\textbf{40} & 36                           & 38                           & 38                           \\
DeepSeek V3            & \cellcolor{green!20}\textbf{40} & \cellcolor{green!20}\textbf{40} & \cellcolor{green!20}\textbf{40} & 38                           \\
DeepSeek R1            & 35                           & 47                           & \cellcolor{green!20}\textbf{50} & 45                           \\
Claude 3.7 Sonnet      & 36                           & 40                           & 40                           & \cellcolor{green!20}\textbf{45} \\
Gemini 2.5 Flash       & \cellcolor{green!20}\textbf{50} & 47                           & \cellcolor{green!20}\textbf{50} & \cellcolor{green!20}\textbf{50} \\
Qwen2.5-7B-Instruct    & 13                           & 15                           & \cellcolor{green!20}\textbf{18} & 13                           \\
CodeLlama-7B           & 4                            & \cellcolor{green!20}\textbf{11} & \cellcolor{green!20}\textbf{11} & 9                            \\
CodeLlama-34B-Instruct & 15                           & \cellcolor{green!20}\textbf{20} & \cellcolor{green!20}\textbf{20} & 13                           \\ \bottomrule
\end{tabular}
\end{table}

\noindent
\textbf{$\bullet$ \black{Batch Size Settings.}}
\black{On the other hand, a batch size of 5 may constrain the diversity of outputs for certain LLMs with higher temperature. To further examine the impact of batch size, we selected two representative models with different optimal temperature settings: GPT-4.1 (which achieves its best Pass@5 at temperature 0.1) and Claude 3.7 (which achieves its best Pass@5 at temperature 1.0). We then increased the batch size from 5 to 100. Table \ref{tab:batch-size100}} presents the comparison. (Due to computational constraints, the 100 batch size experiments were limited to GPT-4.1 and Claude 3.7, and not conducted for the other models.) For GPT-4.1, expanding the batch size had virtually no effect on Pass@1, while Pass@5 improved by 1\%. For Claude 3.7, batch size expansion led to a 1\% increase in Pass@1, but a 2\% decrease in Pass@5. These findings indicate that even under high temperature and large batch size, the overall performance remains virtually the same. Considering computational resources, a batch size of 5 is sufficient to evaluate the performance of LLMs. Therefore, it is recommended that users adopt a batch size of 5 for experimentation.}

\begin{table}[htbp]
\centering
\caption{GPT and Claude Pass@K in batch size 5 and 100}
\label{tab:batch-size100}
\begin{tabular}{lcc}
\toprule
\multirow{2}{*}{\textbf{Model}} & \multicolumn{2}{c}{\textbf{Pass@K}}           \\ \cline{2-3} 
                                & \multicolumn{1}{l}{1} & \multicolumn{1}{l}{5} \\ \midrule
GPT-4.1 (5)                  &  34                            & 38                       \\                   GPT-4.1 (100)                        & 34                    & 39                    \\
Claude 3.7 Sonnet (5)               &32                         & 45    \\
Claude 3.7 Sonnet (100)               & 33                     & 43 \\ \bottomrule
\end{tabular}
\end{table}

\begin{tcolorbox}[size=title,rightrule=1mm, leftrule=1mm, toprule=0mm, bottomrule=0mm, arc=0pt,colback=gray!5,colframe=bleudefrance!75!black,breakable]
\textbf{Answer to RQ4:} Overall, these results suggest that a batch size of 5 combined with a temperature around 0.8 is sufficient for a reliable evaluation in QuanBench. Larger batch sizes offer negligible benefits while incurring significantly higher computational costs.
\end{tcolorbox}

\section{Threats to Validity}\label{sec: threats}

While \textit{QuanBench} includes a diverse quantum programming tasks, several factors may affect the generality of our findings. First, the benchmark is currently limited to the Qiskit framework. Although Qiskit is widely used, models may exhibit different performance when generating code for alternative frameworks such as Cirq or PennyLane. Second, the benchmark focuses primarily on algorithm-level tasks; low-level hardware calibration, pulse-level control, or error mitigation techniques are not included. Third, while we standardize the prompt format and inference parameters, model performance may still vary under alternative prompting strategies, sampling temperatures, or decoding methods. Fourth, although QuanBench includes multiple task categories, the number of quantum machine learning tasks remains limited due to the scarcity of publicly available implementations. These factors may affect the absolute performance reported, although the relative performance trends observed across the models are likely to remain stable. Future work will address these limitations by expanding the task set and supporting additional quantum programming paradigms.

\section{Related Work}\label{sec:relatedwork}

\subsection{Benchmarks for Classical Code Generation}

\deleted{Large language models (LLMs)} \black{LLMs} have demonstrated strong capabilities in code generation in various programming domains~\cite{zheng2023survey}. One of the most influential benchmarks in this area is HumanEval~\cite{chen2021evaluating}, which contains 164 handwritten Python programming problems described via natural language docstrings. It is designed to evaluate the ability of LLMs to synthesize function-level code from natural language specifications.

Following HumanEval, MBPP~\cite{austin2021program} expanded the scope with approximately 1,000 Python tasks created through crowd-sourcing. These problems further assess model performance in synthesizing short programs from descriptive natural language, offering a broader evaluation of generalization and task diversity.

Several domain-specific benchmarks have also been introduced to examine LLM performance in specialized contexts. DS-1000~\cite{lai2023ds} focuses on data science code generation, collecting real-world tasks from StackOverflow in seven popular Python libraries. RMCBench~\cite{chen2024rmcbench} evaluates LLM robustness against adversarial and malicious prompts, assessing model behavior in security-critical scenarios and prompt-level perturbations.

\subsection{Benchmarks for Quantum Code Generation}
Quantum programming is an emerging field, but developing quantum algorithms remains a significant challenge due to the abstract nature of quantum logic and the constraints imposed by circuit design. To evaluate LLMs in this domain, Qiskit HumanEval~\cite{vishwakarma2024qiskit} was proposed as a curated dataset of Qiskit-based tasks designed to test models’ familiarity with quantum programming syntax and APIs. However, this benchmark primarily measures adaptation to the Qiskit syntax, rather than deeper algorithmic reasoning or quantum semantic correctness.

To address this gap, we propose QuanBench, a benchmark specifically designed to evaluate LLMs’ ability to generate quantum algorithm implementations that are both syntactically correct and semantically faithful. \black{To better illustrate the distinction between Qiskit HumanEval and QuanBench, we give two explicit examples. Example~\ref{lst:qh} shows a Qiskit HumanEval docstring, which focuses on API usage: the task specifies adding Hadamard gates, inserting a delay, and returning the circuit, emphasizing step-by-step API compliance. Example~\ref{lst:qb} shows a QuanBench docstring that defines a task in terms of a quantum algorithm (e.g., Bernstein-Vazirani) with additional hints such as qubit ordering. This highlights QuanBench’s focus on algorithmic reasoning and semantic correctness, rather than low-level API calls.}

\subsection{Refined LLM-Generated Quantum Programs}
Current benchmarks have highlighted the limitations of LLMs in generating quantum code. To mitigate this gap, Qiskit HumanEval \cite{vishwakarma2024qiskit} introduced a training dataset and fine-tuned LLMs. In parallel, QSpark \cite{kheiri2025qspark} fine-tuned the Qwen2.5-Coder-32B model using two reinforcement learning methods and evaluated it on Qiskit HumanEval. However, the performance of these fine-tuned models remains limited, which underscores the need for more effective instruction tuning strategies.



\begin{example}[ht]
\begin{lstlisting}[language=Python]
from qiskit import QuantumCircuit

def quantum_circuit_with_delay():
    """
    Create a one-qubit quantum circuit, 
    apply hadamard gate, then add a delay 
    of 100 and then again apply hadamard 
    gate and return the circuit.
    """
    qc = QuantumCircuit(1)
    qc.h(0)
    delay_duration = 100
    qc.delay(delay_duration, 0, unit="dt")
    qc.h(0)
    return qc
\end{lstlisting}
\caption{An example of Qiskit HumanEval benchmark}
\centering
\label{lst:qh}
\end{example}

\begin{example}
\begin{lstlisting}[language=Python]
from qiskit import QuantumCircuit

def Bernstein_Vazirani_011()->QuantumCircuit: 
    """
    Implement the Bernstein Vazirani 
    algorithm for a 3-bit hidden string 
    a = '011', qubits are ordered from 
    right to left (little-endian).
    Return the QuantumCircuit after measure.
    """
    qc = QuantumCircuit(4,3)
    qc.x(3)
    qc.h([0,1,2,3])
    qc.cx(1,3)
    qc.cx(2,3)
    qc.h([0,1,2])
    qc.measure([0,1,2],[0,1,2])
    return qc
\end{lstlisting}
\caption{An example of QuanBench benchmark}
\centering
\label{lst:qb}
\end{example}

\section{Conclusion and Future Work}\label{sec:conclusion}

In this work, we presented QuanBench, a comprehensive benchmark to evaluate the ability of \deleted{large language models (LLMs)}\black{LLMs} to generate code for quantum algorithms. Based on the widely adopted Qiskit framework, QuanBench comprises 44 curated programming tasks spanning diverse algorithmic categories, including Grover's search, the quantum Fourier transform, state preparation, and variational circuits. Each task is paired with an executable canonical solution and evaluated using both functional correctness (via Pass@K) and semantic equivalence (via Process Fidelity).

Our empirical evaluation across multiple state-of-the-art LLMs shows that while models occasionally generate correct quantum programs, their overall performance remains limited. Pass@1 accuracy does not exceed 40\%, and the Process Fidelity scores reveal substantial semantic deviations from canonical solutions. Further error analysis identifies common failure modes, including outdated API usage, structural inconsistencies, and semantic misinterpretation of algorithmic logic.

In particular, models such as DeepSeek R1 and Gemini 2.5 demonstrate relatively higher task coverage and fidelity, suggesting that certain LLMs exhibit emerging strengths in quantum domains. However, a significant performance gap remains, particularly for tasks requiring deeper quantum reasoning and algorithmic understanding.

QuanBench provides a foundation for a systematic and fine-grained evaluation of quantum code generation. Future work includes extending the benchmark to additional frameworks (e.g., Cirq, PennyLane) and investigating techniques such as prompt engineering, fine-tuning, and retrieval-augmented generation to improve performance on quantum tasks.

\section*{Data Availability}
QuanBench used in this paper is publicly available in \href{https://github.com/GuoXiaoYu1125/Quanbench}{https://github.com/GuoXiaoYu1125/Quanbench}.

\section*{Acknowledgments}
This work was supported by JSPS KAKENHI Grant No. JP24K02920, JP23K28062, JP24K14908, and JST BOOST Grant No. JPMJBS2406.

\newpage
\bibliographystyle{IEEEtran}
\bibliography{qse-bibliography}
\end{document}